# COARSE BIFURCATION STUDIES OF BUBBLE FLOW MICROSCOPIC SIMULATIONS

C. Theodoropoulos, K. Sankaranarayanan, S. Sundaresan, I.G. Kevrekidis
Department of Chemical Engineering, Princeton University, Princeton, NJ 08544, USA

**ABSTRACT**
The parametric behavior of regular periodic arrays of rising bubbles is investigated with the aid of 2-dimensional BGK Lattice-Boltzmann (LB) simulators. The Recursive Projection Method is implemented and coupled to the LB simulators, accelerating their convergence towards what we term *coarse* steady states. Efficient stability/bifurcation analysis is performed by computing the leading eigenvalues/eigenvectors of the coarse time stepper. Our approach constitutes the basis for system-level analysis of processes modeled through microscopic simulations.

**INTRODUCTION**
Multiphase flow chemical reactors are central to chemical engineering; they are vital in the production of a variety of chemicals where economy of scale is a driving factor. The hydrodynamic behavior of such gas-liquid and gas-liquid-solid reactors strongly affects their performance and scale-up. In particular, flow details at the level of individual bubbles affect the behavior of the overall system through phase interaction terms and effective stress tensors. It is therefore important both from a scientific and a technological point of view to assess the impact of such terms on the macroscopic hydrodynamics of bubbly flows [1,2].

In this work we use the LB method, based on kinetic theory, to simulate flow characteristics of individual bubbles [3,4]. Our objective is to expand the capabilities of the microscopic LB time-steppers, enabling them to perform *system level* tasks, such as stability analysis, continuation and bifurcation computations. Such tasks are currently inaccessible to microscopic simulators. *Coarse* (averaged) steady states of the system correspond to stationary profiles of moments of distributions from microscopic simulations. The parameter-dependent behavior of a system is traditionally analyzed by first obtaining a coarse model (e.g. a mean field PDE) and then performing bifurcation/stability analysis of this PDE. Here we propose an alternative methodology, building on the time-stepper-based approach to PDE bifurcation analysis. In particular, we adapt the Recursive Projection Method (RPM) of Shroff & Keller [5] to accelerate the convergence of the LB simulation to stable, and even to unstable *coarse* stationary (macroscopically steady) states. This implements coarse bifurcation analysis using the misroscopic evolution rules directly, and avoiding the intermediate step of construction of a coarse, mean-field PDE [6]. Such an approach is essential when a macroscopic description of the system is unavailable in closed form.

Two-dimensional parametric studies of individual bubbles using the LB method are performed using periodic boundary conditions for the computational domain, thus corresponding to regular arrays of rising bubbles. Transitions to oscillatory patterns and formation of unstable wakes [2,7,8] are investigated. Our bifurcation parameter corresponds to variation of the average density difference between the gas and the liquid phase. The combination of LB and RPM allows the detection of coarse Hopf bifurcations and the corresponding coarse spectra (eigenvalues and eigenvectors).

## LATTICE BOLTZMANN SIMULATIONS

Simulations of a single bubble rising in 2-D doubly periodic domains are performed employing a single species BGK-LB model [9] with a non-ideal equation of state. The governing evolution rule for this model, derived from the continuum BGK-Boltzmann distribution function for a single particle $f(x,\xi,t)$ in position ($x$) and velocity ($\xi$) phase space and time t, is given by the dimensionless explicit formula:

$$f_i(\mathbf{x}+\xi_i, t+1) - f_i(\mathbf{x},t) = -\frac{f_i(\mathbf{x},t) - f_i^{eq}(\mathbf{x},t)}{\tau} \quad (1)$$

Here $\tau$ is the relaxation time, related to the kinematic viscosity $\nu$, $\nu=\tau-1/2$; $f_i$'s are related to $f(x,\xi,t)$ through: $f_i = (w_i/w(\xi_i))f(x,\xi,t)$ where $w_i(\xi)$ are weights, $w(\xi_i)$ weight functions: $w(\xi_i)=(1/2\pi)^{D/2}\exp(-\xi_i\xi_i/2)$ (D=2, the dimension of the problem) and $\xi_i$'s the abscissas of the quadrature in velocity space as given in [10,11] for a square lattice in 2 dimensions:

$$\xi_i = \begin{cases} (0,0) & i=0 \\ (\cos\phi_i, \sin\phi_i)\sqrt{3}, \phi_i=(i-1)\pi/2 & i=1-4 \\ (\cos\phi_i, \sin\phi_i)\sqrt{6}, \phi_i=(i-5)\pi/2+\pi/4 & i=5-8 \end{cases} \quad w_i = \begin{cases} 4/9, i=0 \\ 1/9, i=1-4 \\ 1/36, i=5-8 \end{cases} \quad (2)$$

It has been shown [10] that a second order Hermite polynomial approximation for the equilibrium distribution $f^{eq}$, is sufficient for isothermal flow calculations:

$$f_i^{eq} = w_i n[1 + \gamma_i + \xi_i \cdot (\mathbf{u}+\tau\mathbf{a}) + 1/2(\xi \cdot (\mathbf{u}+\tau\mathbf{a}))^2 - (\mathbf{u}+\tau\mathbf{a})\cdot(u+\tau\mathbf{a})/2] \quad (3)$$

where the local density n and the momentum n**u** (**u** being the local velocity vector) on a lattice node can be uniquely computed from successive moments of $f_i$:

$$n = \sum_{i=0}^{8} f_i \qquad n\mathbf{u} = \sum_{i=0}^{8} f_i \xi_i \quad (4)$$

The specific force **a** consists of the external specific force $\mathbf{a}_{ext}$ and that due to particle interactions, $\mathbf{a}_{int}$: $\mathbf{a}=\mathbf{a}_{ext}+\mathbf{a}_{int}$. The external force which causes the bubble to rise is set to $\mathbf{a}_{ext}=\mathbf{g}(1 - <n>/n)$ <n> being the average density of the mixture in the entire (periodic) domain and $\mathbf{g}$ the acceleration due to gravity. The internal force according to [12,13] is set to $\mathbf{a}_{int}=G\sum\psi(x+\xi_i)\psi(x)\xi_i$, $i=0,...8$, where G denotes the Green's function (here set to 2.0) and $\psi$ the interaction potential, chosen according to [12]: $\psi=1-\exp(-n)$. The equation of state for the system is [12]: $P = n\theta - G/2\psi^2$; P is the pressure and $\theta$ the dimensionless temperature (equal to 1 for an isothermal system).

The LB simulations were performed in periodic domains consisting of 128x128 lattice nodes. The area fraction of the bubble was ~5.8% and $\nu$ was set to 0.5. The important dimensionless groups encountered in gas-liquid bubbly flows are the Reynolds number (Re=$ud/\nu$), the Morton number (Mo=$g\rho^2\Delta\rho\nu^4/\sigma^3$) and the Eötvös number (Eo=$g\Delta\rho d^2/\sigma$); $\rho$ being the density of the liquid phase, $d$ the effective diameter and $u$ the slip velocity of the bubble, $\sigma$ the interfacial tension and $\Delta\rho$ the density difference between the liquid and the gas. The parameter varied here was $g$, which, through the Mo and Eo numbers, corresponds to a change in $\Delta\rho$. At low values of $g$ ($g$=0.00015 corresponding to Eo=3.0, Mo=0.0019; here Eo≈19994.3g and Mo≈12.6g) the bubble is almost spherical and rises vertically in the periodic domain. In fig.1a three density snapshots are shown; dark grey color denotes low density (gas), light grey high density (liquid) and white the interface between the two phases. When $g$ increases, a wake forms causing bubble shape deformation. Beyond a threshold value of $g$ the wake starts shedding, and the shape of the rising bubble starts oscillating, clearly suggesting a Hopf bifurcation in $g$. Fig.1b shows snapshots of an oscillating rising bubble for $g$=0.0006.

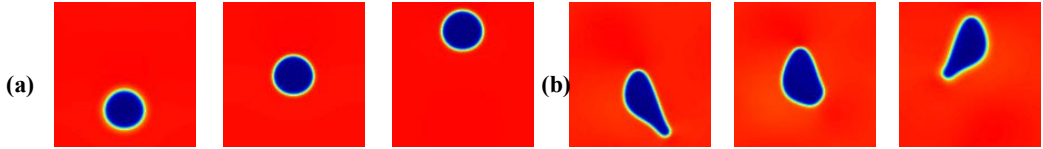

**Figure 1: (a)** Density snapshots of a steadily rising bubble for g=0.00015 (Eo=3, Mo=0.0019). (b) Density snapshots of an oscillating rising bubble for *g*=0.0006 (Eo=12, Mo=0.0075).

A phase space projection of the oscillation is shown in fig. 2; its symmetric shape underscores the spatiotemporal symmetry of the bubble oscillations. Characterizing the long-term behavior through LB requires extensive simulations in time and no systematic way exists to pinpoint the transition parameter value. Also, beyond the Hopf bifurcation, the LB simulator cannot find the unstable steady states. Adapting RPM and building it around the LB code, efficiently converts the LB timestepper to a steady state continuation code for coarse bifurcation studies.

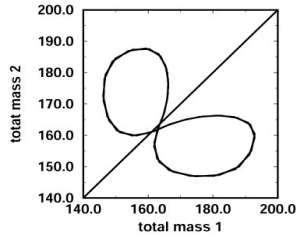

**Figure 2:** Phase space plot of the "total mass 1", five lattice lines left of the vertical centerline vs. "total mass 2", five lattice lines right of the vertical centerline for g=0.0006.

## THE RECURSIVE PROJECTION METHOD

RPM constitutes a software superstructure built around an existing time-stepping code (here the LB simulator). Steady states of a system are found as fixed points of the time-stepper. Consecutive calls to the time-stepper for several nearby initial conditions (IC's) and short periods of time constitute the backbone of the method. RPM treats the time-integration routine as a black box and, exploiting the existence of a gap between the strongly stable and the slow eigenmodes of the problem (separation of time-scales) adaptively identifies a low-dimensional invariant subspace **P** along which the convergence is the slowest, even slightly unstable. Integrations towards the steady state (Picard iterations) are then substituted by Newton iterations in **P**, using a small approximate numerical Jacobian, and Picard iterations in its orthogonal complement **Q**. The method sidesteps, therefore, the construction and inversion of large Jacobians and is intended for the efficient bifurcation analysis of large-scale systems.

As discussed in the introduction, we have adapted RPM for use in conjunction with *microscopic* evolution rules (the LB integrator) for *coarse* bifurcation studies without first obtaining a corresponding PDE-based model that evolves density and momentum fields for the two phases (e.g.[2]). Our methodology consists of evolving coarse profiles using only microscopic rules by: (i) translating (lifting) a coarse IC (density and momentum fields of the system) to a set of consistent microscopic ones; (ii) evolving the realizations using microscopic rules (here LB) for some time interval (the reporting horizon, T); (iii) appropriately averaging the results over fine space and/or fine time and/or number of realizations to obtain a coarse time-T map. The lifting step to microscopic IC's (here the $f_i$'s –equation 1) is not unique; here we were guided by the distribution of equation 3. Further discussions are given in [6].

## RESULTS AND DISCUSSION

A RPM-LB continuation algorithm was developed in order to study the coarse bifurcation behavior of the rising bubble. The continuation parameter, $g$, ranged from 0.0001 to 0.007; the reporting horizon was T=1200 LB steps. The RPM-LB code converged to the coarse steady states faster than direct LB, and found even unstable ones. The dimension, m, of **P** varied from m=2 far from the instability, to 6 near the Hopf point. The coarse eigenvalues of **P** indicated a Hopf bifurcation at $g \approx 0.000475$. A detailed bifurcation diagram is depicted in fig. 3a. The solid (broken) line denotes stable (unstable) steady states, with representative bubble shapes shown in the insets. The coarse eigenspectrum at g=0.0005 (beyond the Hopf, with a pair of complex eigenvalues outside the unit circle) is depicted in fig. 3b.

Certain mathematical issues arise in the implementation of the LB-RPM procedure: The LB method implicitly conserves total mass and momentum. The construction of the small numerical Jacobian, therefore, by perturbation of coarse initial profiles, should respect mass and momentum conservation. Furthermore, the system is translationally invariant in both directions. Pinning conditions need to be implemented to isolate solutions. Both issues were efficiently resolved by implementing the coarse profiles in Fourier space. Conservation was ensured by keeping the $0^{th}$ Fourier mode constant both for densities and momenta. Our Fourier space pinning (see [15] for details) was implemented through the use of templates [14]. Partial support: a ΠENEΔ grant (Greece); and AFOSR (Dynamics and Control, USA).

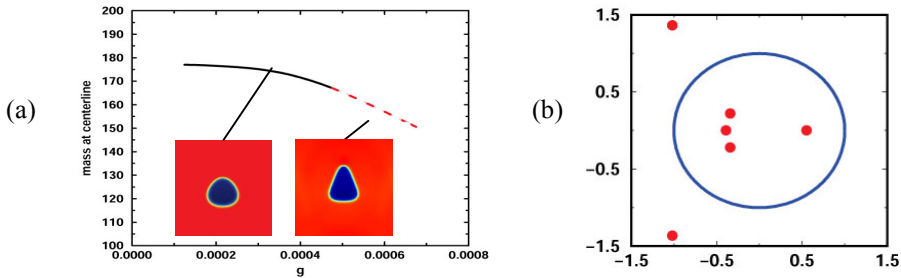

**Figure 3**: (a) Coarse bifurcation diagram. Total mass at vertical centerline is plotted vs. g. Solid (boken) line: stable (unstable) steady states. Insets: steady bubble shapes at g=0.0003 and g=0.0006. (b) Coarse eigenspectrum at g=0.0005 corresponding to Eo=10 and Mo=0.0063.